\documentclass[a4paper,12pt]{article}
\usepackage{ragged2e}
\usepackage{biblatex} 
\usepackage[english]{babel}
\usepackage[nottoc]{tocbibind}
\usepackage[colorlinks=true,citecolor=blue]{hyperref}
\usepackage{atbegshi}
\AtBeginDocument{\AtBeginShipoutNext{\AtBeginShipoutDiscard}}
\usepackage{amsmath}
\usepackage{multirow}
\usepackage{multicol}
\usepackage{diagbox}
\newcommand\authormark[1]{\textsuperscript{#1}}
\usepackage{pgfplots}
\pgfplotsset{width=10cm,compat=1.9}
\usepackage{pdflscape}
\usepackage{graphbox}
\usepackage{mwe}
\usepackage{geometry}
\usepackage{graphicx}
\graphicspath{ {./images/} }

\title{\textbf{Using ResNet to Utilize 4-class T2-FLAIR Slice Classification Based on the Cholinergic Pathways Hyperintensities Scale for Pathological Aging}}
\author{
Wei-Chun Kevin Tsai\authormark{1,†}, Yi-Chien Liu\authormark{2}, Ming-Chun Yu\authormark{2},\\
Chia-Ju Chou\authormark{2}, Sui-Hing Yan\authormark{3}, Yang-Teng Fan\authormark{4},\\
Yan-Hsiang Huang\authormark{2}, Yen-Ling Chiu\authormark{5},\\
Yi-Fang Chuang\authormark{6,7}, Ran-Zan Wang\authormark{1}, and Yao-Chia Shih\authormark{4}
\vspace{5mm}
\\
\authormark{1}Computer Science and Engineering, Yuan Ze University,\\Taoyuan City, Taiwan\\
\authormark{2}Neurology, Cardinal Tien Hospital,\\New Taipei City, Taiwan\\
\authormark{3}Neurology, Far Eastern Memorial Hospital,\\New Taipei City, Taiwan\\
\authormark{4}Graduate Institute of Medicine, Yuan Ze University,\\Taoyuan City, Taiwan\\
\authormark{5}Medical Research, Far Eastern Memorial Hospital,\\New Taipei City, Taiwan\\
\authormark{6}Psychiatry, Far Eastern Memorial Hospital,\\New Taipei City, Taiwan\\
\authormark{7}Institute of Public Health, National Yang Ming Chiao Tung University,\\Taipei City, Taiwan\\
}
\date{}

\begin{document}
\justify

\setcounter{page}{0}

\maketitle

\clearpage

\begin{center}
    \subsection*{Synopsis}    
\end{center}
\subsubsection*{Motivation:}
Cholinergic Pathways Hyperintensities Scale (CHIPS) is a visual rating scale to evaluate the burden of cholinergic white matter hyperintensities in T2-FLAIR image, indicating the severity of dementia. However, it is still time-consuming to screen slices throughout the whole brain to choose 4 specific slices for rating.

\subsubsection*{Goal(s):}
To develop a deep-learning-based model to automatically select 4 slices specific to CHIPS.

\subsubsection*{Approach:}
We used ADNI T2-FLAIR dataset (N=150) to train a 4-class slice classification model (BSCA) utilized by ResNet, and a local dataset (N=30) to
test its performance.

\subsubsection*{Results:}
Our model achieved the accuracy of 99.82\% and F1-score of 99.83\%.

\subsubsection*{Impact:}
BSCA can be an automatic screening tool to efficiently provide 4 specific T2-FLAIR slices covering the white matter landmarks along the cholinergic pathways for clinicians to help evaluate whether patients have the high risk to develop clinical dementia.

\section{Introduction}
T2-weighted fluid-attenuated-inversion-recovery (T2-FLAIR) magnetic resonance Imaging (MRI) is clinically used to detect and visualize white matter hyperintensities (WMH) as brain lesions and utilized in routine clinical practice. Normally, clinicians will review T2-FLAIR axial slices throughout the whole brain, and then visually rate the Fazekas scale score \cite{doi:10.2214/ajr.149.2.351} to provide an overall impression of WHM burden. However, the Fazekas scale lacks the specificity to evaluate the severity of functional declines in dementia, and limits it clinical utility. The recent study done by our team adopted another visual rating scale as known as the Cholinergic Pathways Hyperintensities Scale (CHIPS)\cite{Bocti2005-zh}, which is specific to the severity of WMH burden in the cholinergic pathway associated with the Clinical Dementia Rating scale Sum of Boxes (CDR-SB), reflecting clinical dementia severity in APOE e4 carriers \cite{10.3389/fneur.2023.1100322}. It implies that CHIPS might have higher diagnostic value than Fazekas scale. CHIPS evaluation is based on only four slices, separately covering (1) low external capsule, (2) higher external capsule and anterior cingulate gyrus, (3) corona radiata and posterior cingulate gyrus, and (4) centrum semiovale, facilitating diagnostic efficiency and specificity. However, it still requires experienced clinicians to visually recognize and classify these four slices. Therefore, the present study aimed to develop an automatic slice selection model to assist in classifying the aforementioned four axial slices in T2-FLAIR images to provide a quicker clinical screening for APOE e4 carriers with clinical dementia.

\section{Methods and Materials}
\begin{landscape}
\begin{figure}
    \centering
    \includegraphics[width=230mm]{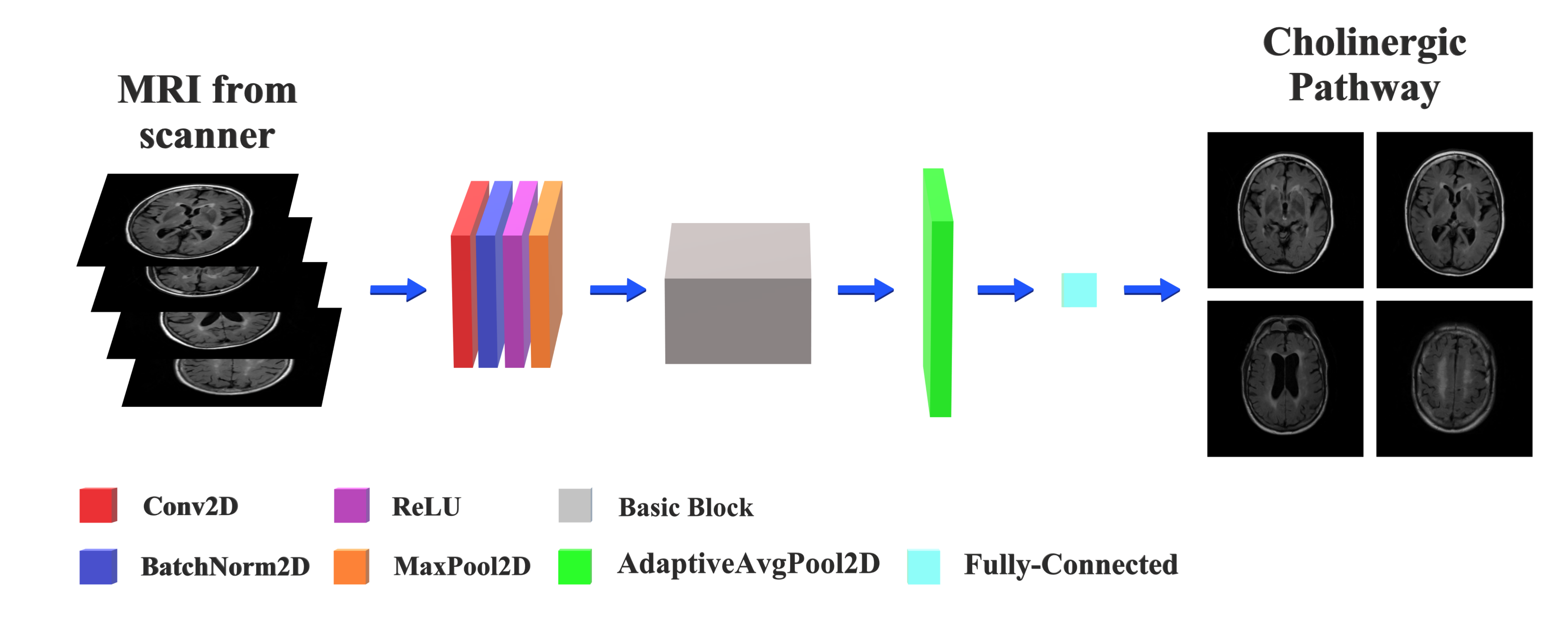}
    \caption{Residual network architecture}
    \label{fig:nn-network-preview}
\end{figure}
\end{landscape}

Here we introduced a Brain Slice Classification Algorithm (BSCA) which was built up based on a convolutional neural network, specifically utilized by the residual network (ResNet)\cite{he2015deep}. This approach enables a deep-learning scheme to automatically choose specific brain MRI slices without additional image annotations. Figure \ref{fig:nn-network-preview} and Table \ref{tab:resnet_table}  illustrates the whole BSCA architecture. Multi-center T2-FLAIR datasets were obtained from two sources: Alzheimer’s Disease Neuroimaging Initiative (ADNI) (with the following acquisition parameters using 3T MRI scanners across vendors: TR/TE/TI=9000-11000/90-154/2250-2500 ms, pixel spacing=0.8594 mm, slice thickness=5 mm) \cite{Petersen2009-mw} and Taiwan Precision Medicine Initiative in Cognition (TPMIC, with the following acquisition parameters using two 3T MRI scanners [Skyra, Siemens, Erlangen, Germany]: TR/TE/TI=8000/85/2370 ms, pixel spacing=0.6875 mm, slice thickness=6.5 mm) contributed by two local hospitals. T2-FLAIR data (N=150) of ADNI dataset \cite{Petersen2009-mw} comprising of patients with early/late stages mild cognitive impairment (MCI) and Alzherimer’s dementia (AD) were used as a training dataset, whereas those of TPMIC dataset of MCI and AD (N=30) were used as a testing dataset. BSCA was trained in eight-fold cross-validation based on the slice-level data splitting (4504 slices in total) to classify 4 different slices specific to CHIPS evaluation. The performance of BSCA was assessed using the metrices of accuracy, loss, precision, recall, and F1-score. We adopted Adam as an optimizer to minimize the loss function based on cross-entropy. Note that the specific 4 slices in the training data were labeled by a well-trained graduate student (W.C. Tsai) trained by an experienced neurologist (M.C. Yu).

\clearpage

\begin{landscape}
    \begin{table}[h]
        \centering
        \begin{tabular}{|l|l|l|c|}
        \hline
            Layer & Operation & Output Shape & Parameters \\
        \hline
            Input & Input Full-Color Image & 3 x 256 x 256 & \\
            Convolutional Layer 1 (Conv1) & Conv2d(3, 64, kernel=7, stride=2, padding=3) & 64 x 128 x 128 & 9,408 \\
            Batch Normalization Layer 1 (BatchNorm1) & BatchNorm2d(64) & 64 x 128 x 128 & 128 \\
            ReLU Activation (ReLU) & ReLU(inplace=True) & 64 x 128 x 128 & \\
            Max Pooling Layer (MaxPool) & MaxPool2d(kernel=3, stride=2, padding=1) & 64 x 64 x 64 & \\
            Layer 1 (Convolutional Blocks) & Two BasicBlocks & 64 x 64 x 64 & \\
            BasicBlock1 & Includes convolution and normalization & 64 x 64 x 64 &\\
            BasicBlock2 & Includes convolution and normalization & 64 x 64 x 64 &\\
            Layer 2 (Convolutional Blocks) & Two BasicBlocks & 128 x 32 x 32 & \\
            BasicBlock1 & Includes convolution and normalization & 128 x 32 x 32 & \\
            BasicBlock2 & Includes convolution and normalization & 128 x 32 x 32 & \\   
            Layer 3 (Convolutional Blocks) & Two BasicBlocks & 256 x 16 x 16 & \\
            BasicBlock1 & Includes convolution and normalization & 256 x 16 x 16 & \\
            BasicBlock2 & Includes convolution and normalization & 256 x 16 x 16 & \\
            Layer 4 (Convolutional Blocks) & Two BasicBlocks & 512 x 8 x 8 & \\
            BasicBlock1 & Includes convolution and normalization & 512 x 8 x 8 & \\
            BasicBlock2 & Includes convolution and normalization & 512 x 8 x 8 & \\
            Adaptive Average Pooling Layer (AdaptiveAvgPool) & AdaptiveAvgPool2d((1, 1)) & 512 x 1 x 1 & \\
            Flatten Layer (Flatten) & Reshape to a 1D tensor & 512 & \\
            Fully Connected Layer & Linear(512, 4 classes) & 4 classes & 205,004 \\
            \hline
        \end{tabular}
        \caption{A table of the residual network architecture applied to classify 4 specific T2-FLAIR MRI slices}
        \label{tab:resnet_table}
    \end{table}
\end{landscape}

\begin{center}
\begin{figure}[h]
\centering
\begin{tikzpicture}
\begin{axis}[
    title={Multi-Metric Performance Comparison},
    xlabel={Epoch},
    ylabel={Metrics Rate},
    xmin=0, xmax=12,
    ymin=0, ymax=1.2,
    xtick={0,2,4,6,8,10,12},
    ytick={0,0.2,0.4,0.6,0.8,1.0,1.2},
    legend pos=south west,
    ymajorgrids=true,
    grid=both,
]

\addplot[
    color=blue,
    mark=square,
    ]
    coordinates {
        (1,0.5078)
        (2,0.7593)
        (3,0.8629)
        (4,0.9336)
        (5,0.9589)
        (6,0.9716)
        (7,0.9803)
        (8,0.9791)
        (9,0.9904)
        (10,0.9986)
    };

\addplot[
    color=red,
    mark=diamond
    ]
    coordinates {
        (1,1.1763)
        (2,0.5741)
        (3,0.3399)
        (4,0.1944)
        (5,0.1287)
        (6,0.0830)
        (7,0.0581)
        (8,0.0686)
        (9,0.0326)
        (10,0.0196)
    };

\addplot[
    color=green,
    mark=triangle,
    ]
    coordinates {
        (1,0.4961)
        (2,0.7566)
        (3,0.8616)
        (4,0.9336)
        (5,0.9588)
        (6,0.9718)
        (7,0.9803)
        (8,0.9791)
        (9,0.9904)
        (10,0.9981)
    };

\addplot[
    color=orange,
    mark=x,
    ]
    coordinates {
        (1,0.5078)
        (2,0.7593)
        (3,0.8629)
        (4,0.9336)
        (5,0.9589)
        (6,0.9716)
        (7,0.9803)
        (8,0.9791)
        (9,0.9904)
        (10,0.9986)
    };

\addplot[
    color=brown,
    mark=*,
    ]
    coordinates {
        (1,0.4986)
        (2,0.7574)
        (3,0.8619)
        (4,0.9336)
        (5,0.9588)
        (6,0.9717)
        (7,0.9803)
        (8,0.9791)
        (9,0.9904)
        (10,0.9983)
    };
    \legend{Accuracy, Loss, Precision, Recall, F1-Score}

\end{axis}
\end{tikzpicture}
\caption{The model performance metrics}
\label{tik:train-validation}
\end{figure}
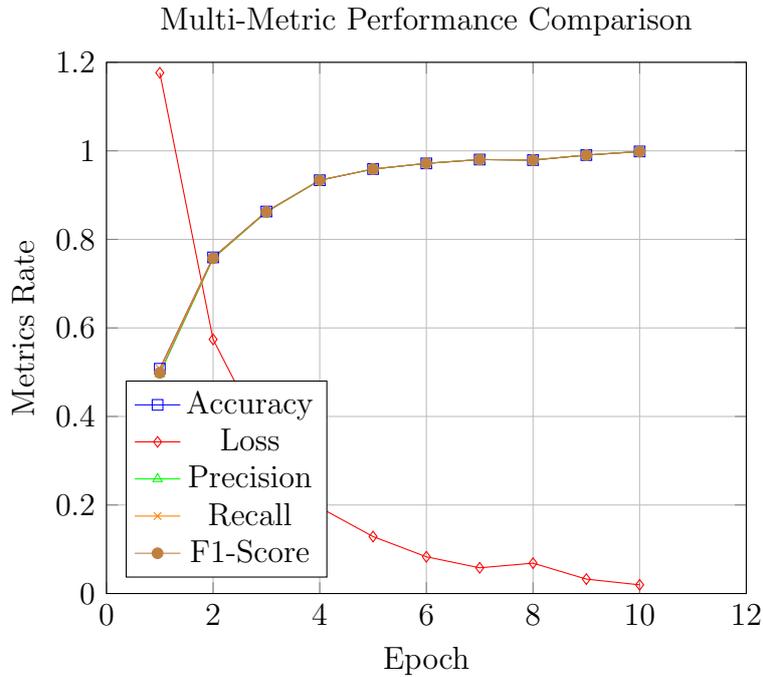
\end{center}

\section{Results}
After training BSCA with learning rate=0.05, batch size=8, and 10 epochs, BSCA reached the best performance of 4-class slice classification, with the accuracy of 99.82\%, precision of 99.81\%, recall of 99.86\%, and F1-score of 99.83\% (Figure \ref{tik:train-validation}). Table \ref{tab:result-table} demonstrates the output of BSCA, and it successfully selected 4 different classes of T2-FLAIR slices corresponding to 4 CHIPS anatomical slices from the 4 patients randomly chosen from TPMIC dataset.

\section{Discussion and Conclusion}
We demonstrated that BSCA utilized by ResNet\cite{he2015deep} can be an automatic screening tool to efficiently provide 4 specific T2-FLAIR slices covering the white matter landmarks along the cholinergic pathways \cite{Bocti2005-zh, Selden1998-by} for clinicians to evaluate whether patients have the high risk to develop clinical dementia \cite{10.3389/fneur.2023.1100322}. To add clinically diagnostic value to BSCA, future work will apply it as a module to a WMH segmentation tool that can automatically identify the white matter lesions and compute the CHIPS scores to rate the severity of WMH burden and assist in evaluating the disease severity of clinical dementia.

\begin{landscape}
    \begin{table}[h]
        \centering
        \begin{tabular}{|c|c|c|c|c|c|}
            \hline
            \diagbox{Brain Slice ID}{Patient No.} & 1 & 2 & 3 & 4 & Coronal View (E)\\
            \hline
            A & \includegraphics[width=30mm, align=c]{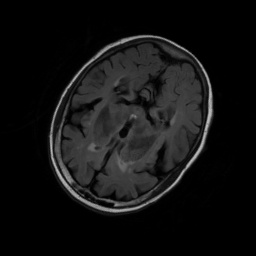} & \includegraphics[width=30mm, align=c]{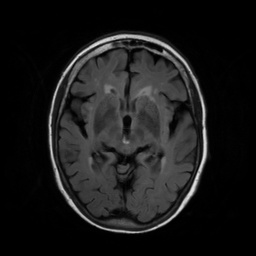} & \includegraphics[width=30mm, align=c]{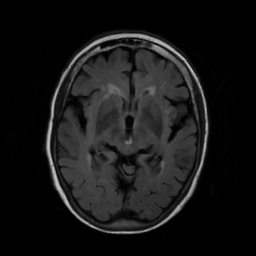} & \includegraphics[width=30mm, align=c]{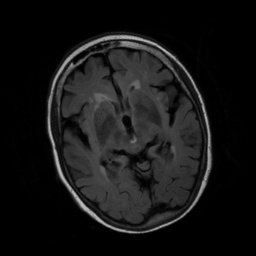} & \includegraphics[width=25mm, align=c]{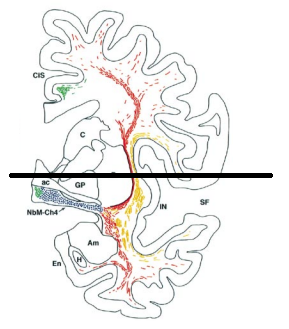} \\
            \hline
            B & \includegraphics[width=30mm, align=c]{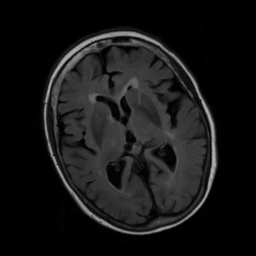} & \includegraphics[width=30mm, align=c]{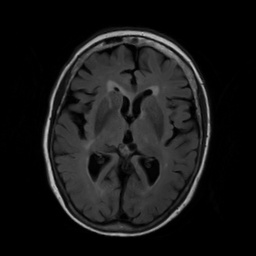} & \includegraphics[width=30mm, align=c]{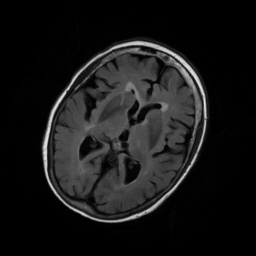} & \includegraphics[width=30mm, align=c]{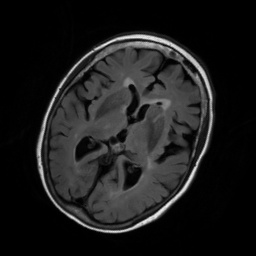} & \includegraphics[width=25mm, align=c]{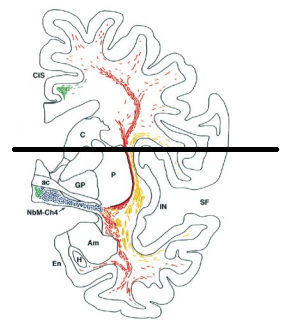} \\
            \hline
            C & \includegraphics[width=30mm, align=c]{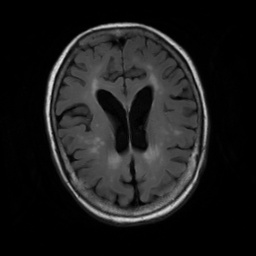} & \includegraphics[width=30mm, align=c]{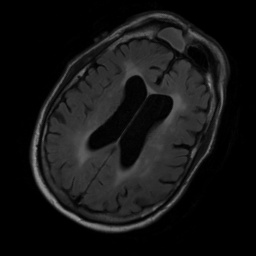} & \includegraphics[width=30mm, align=c]{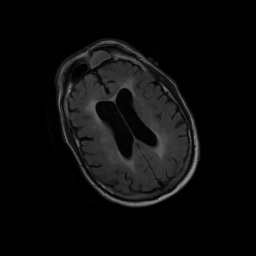} & \includegraphics[width=30mm, align=c]{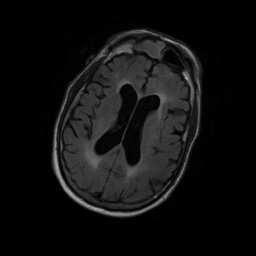} & \includegraphics[width=25mm, align=c]{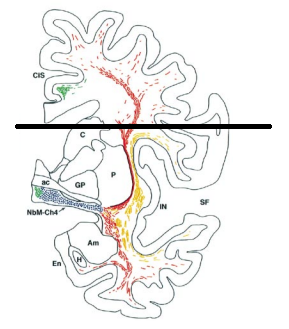} \\
            \hline
            D & \includegraphics[width=30mm, align=c]{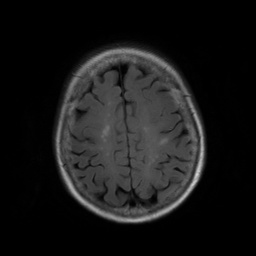} & \includegraphics[width=30mm, align=c]{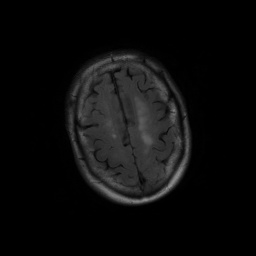} & \includegraphics[width=30mm, align=c]{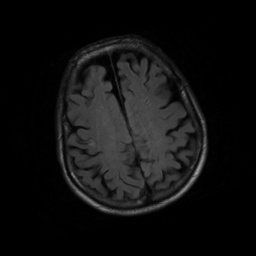} & \includegraphics[width=30mm, align=c]{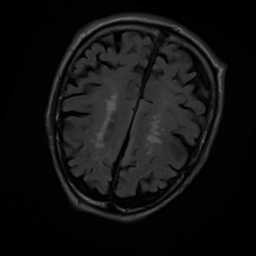} & \includegraphics[width=25mm, align=c]{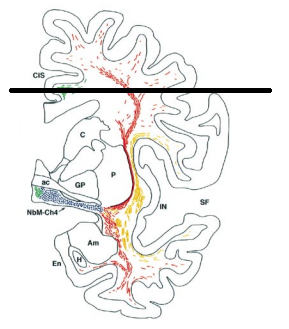} \\
            \hline
        \end{tabular}
        \caption{The testing outcomes of our proposed BSCA model. Four patients from a local T2-FLAIR dataset were randomly selected to test. BSCA is successful to automatically pick up 4 slices respectively corresponding to 4 specific planes from the inferior layer to the posterior layer along the cholinergic pathways shown by the coronal view of a schematic: (A) low external capsule, (B) high external capsule and anterior gyrus, (C) coronal radiata and posterior cingulate gyrus, (D) centrum semiovale; (E) a schematic showing 4 different levels is adopted from Selden et al.\cite{Selden1998-by}}
        \label{tab:result-table}
    \end{table}
\end{landscape}

\section{Acknowledgments}
Funding for this project was obtained from the National Science and Technology Concil of Taiwan (NSTC-110-2321-B418-001) and Cardinal Tien Hospital (CTH-110-2-1-014).

\medskip
\printbibliography
\end{document}